\documentclass[conference]{IEEEtran}
\ifCLASSINFOpdf
   \usepackage[pdftex]{graphicx}
\else
   \usepackage[dvips]{graphicx}
\fi
\usepackage[cmex10]{amsmath}

\usepackage{multirow}
\usepackage{subfig}
\usepackage{epsfig}
\usepackage{amssymb}
\usepackage{comment}

\usepackage{array}
\begin{document}
%
\title{Automatic Document Image Binarization using Bayesian Optimization}




\author{\IEEEauthorblockN{Ekta Vats, Anders Hast and Prashant Singh}
\IEEEauthorblockA{Department of Information Technology\\
Uppsala University, SE-751 05 Uppsala, Sweden \\
Email: ekta.vats@it.uu.se; anders.hast@it.uu.se; prashant.singh@it.uu.se}
}


%


\maketitle

\begin{abstract}

Document image binarization is often a challenging task due to various forms of degradation. Although there exist several binarization techniques in literature, the binarized image is typically sensitive to control parameter settings of the employed technique. This paper presents an automatic document image binarization algorithm to segment the text from heavily degraded document images. The proposed technique uses a two band-pass filtering approach for background noise removal, and Bayesian optimization for automatic hyperparameter selection for optimal results. The effectiveness of the proposed binarization technique is empirically demonstrated on the Document Image Binarization Competition (DIBCO) and the Handwritten Document Image Binarization Competition (H-DIBCO) datasets.

\end{abstract}


%
\IEEEpeerreviewmaketitle

\section{Introduction}
\label{sec:intro}
Document image binarization aims to segment the foreground text in a document from the noisy background during the preprocessing stage of document analysis. Document images commonly suffer from various degradations over time, rendering document image binarization a daunting task. Typically, a document image can be heavily degraded due to ink bleed-through, faded ink, wrinkles, stains, missing data, contrast variation, warping effect, and noise due to lighting variation during document scanning.

Though document image binarization has been extensively studied, thresholding of heavily degraded document images remains a largely unexplored problem due to difficulties in modelling different types of document degradations. The Document Image Binarization Competition (DIBCO), and the Handwritten Document Image Binarization Competition (H-DIBCO), held from 2009 to present aim to address this problem by introducing challenging benchmarking datasets to evaluate the recent advancement in document image binarization. However, competition results so forth indicate a scope for improvement in the binarized image quality.

The performance of binarization techniques significantly depends on the associated control parameter values \cite{howe2013document}, i.e., the hyperparameters. Despite the significance of optimal hyperparameter selection for document image binarization, automatic binarization has still not been sufficiently explored. This paper presents an automatic document image binarization technique that uses two band-pass filters for background noise removal, and Bayesian optimization \cite{shahriari2016taking} for automatic thresholding and hyperparameter selection. The band-pass filtering method uses a high frequency band-pass filter to separate the fine detailed text from the background, and subsequently a low frequency band-pass filter as a mask to remove noise. The parameters of the two band-pass filtering algorithm include a threshold for removing noise, the mask size for blurring the text, and the window size to be set dynamically depending upon the degree of degradation. Bayesian optimization is used to automatically infer the optimal values of these parameters.

The proposed method is simple, robust and fully automated for handling heavily degraded document images. This makes it suitable for use by, e.g., librarians and historians for quick and easy binarization of ancient texts. Since optimum parameter values are selected on-the-fly using Bayesian optimization, the average binarization performance is improved. This is due to each image being assigned its respective ideal combination of hyperparameter values instead of using a global sub-optimal parameter setting for all images. To the best of authors' knowledge, this is the first work in the community that uses Bayesian optimization on binarization algorithm for selecting multiple hyperparameters dynamically for a given input image.


\begin{figure*}
\centering
\includegraphics[width=4.8in]{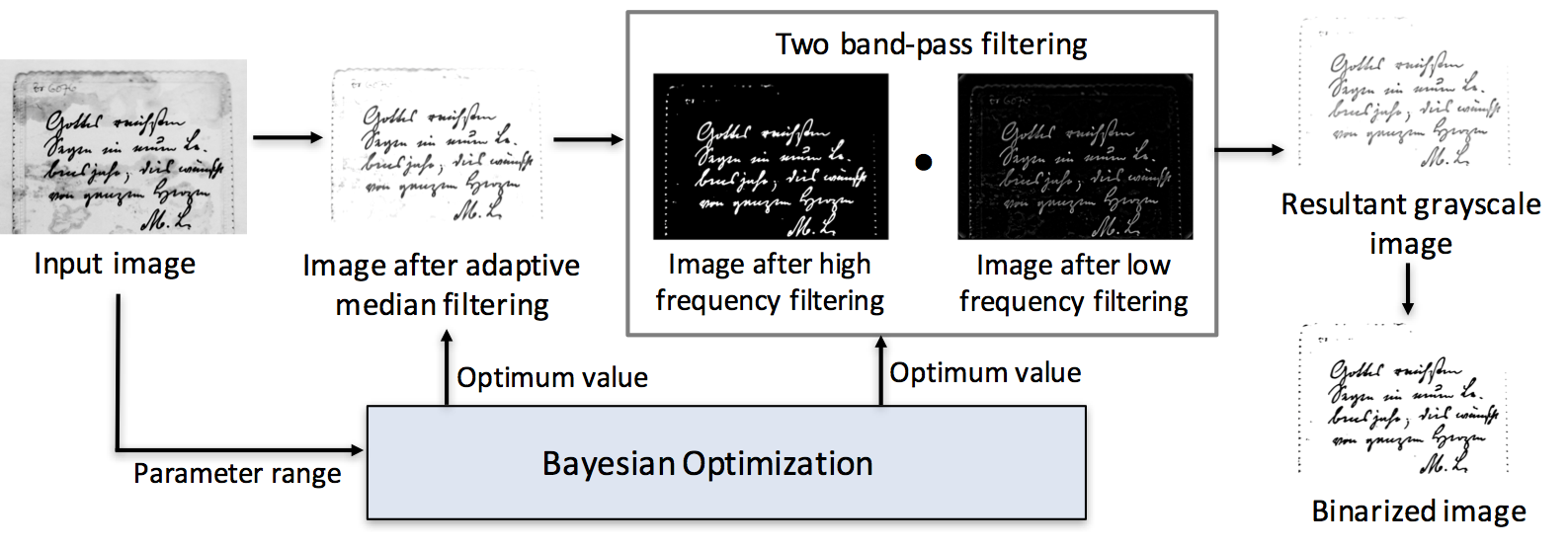}
\caption{The proposed automatic document image binarization framework.}
\label{fig_framework}
\end{figure*}

\section{Document binarization methods}
\label{sec:literature}
Numerous document image binarization techniques have been proposed in literature, and are well-documented as part of the DIBCO reports \cite{gatos2009icdar,pratikakis2010h,gatos2011icdar2,pratikakis2012icfhr,pratikakis2013icdar,ntirogiannis2014icfhr2014,pratikakis2016icfhr2016}. Under normal imaging conditions, a simple global thresholding approach, such as Otsu's method \cite{otsu1975threshold}, suffices for binarization. However, global thresholding is not commonly used for severely degraded document images with varying intensities. Instead, an adaptive thresholding approach that estimates a local threshold for each pixel in a document is popular \cite{bernse1986dynamic,sauvola2000adaptive,niblack1985introduction}. In general, the local threshold is estimated using the mean and standard deviation of the pixels in a document image within a local neighborhood window. However, the prime disadvantage of using adaptive thresholding is that the binarization performance depends upon the control parameters such as the window size, that cannot be accurately determined without any prior knowledge of the text strokes. Moreover, methods such as Niblack's thresholding \cite{niblack1985introduction} commonly introduce additional noise, while Sauvola's thresholding \cite{sauvola2000adaptive} is highly sensitive to contrast variations.

Other popular methods in literature include \cite{lu2010document,gatos2006adaptive,howe2011laplacian,howe2013document,su2010binarization,lelore2011super,su2013robust}. The binarization algorithms of the winners of competitions are sophisticated, and achieve high performance partially due to thresholding, but primarily by modelling the text strokes and the background to enable accurate pixel classification. Lu \textit{et al.} \cite{lu2010document} modelled the background using iterative polynomial smoothing, and a local threshold was selected based on detected text stroke edges. Lelore and Bouchara \cite{lelore2011super} proposed a technique where a coarse threshold is used to partition pixels into ink, background, and unknown groups. The method is based on a double threshold edge detection approach that is capable of detecting fine details along with being robust to noise. However, these methods combine a variety of image related information and domain specific knowledge, and are often complex \cite{su2013robust}. For example, methods proposed in \cite{gatos2006adaptive,su2010binarization} made use of document specific domain knowledge. Gatos \textit{et al.} \cite{gatos2006adaptive} estimated the document background based on the binary image generated using Sauvola's thresholding \cite{sauvola2000adaptive}. Su \textit{et al.} \cite{su2010binarization} used image contrast evaluated based on the local maximum and minimum to find the text stroke edges.

Although there exist several document thresholding methods, automatic selection of optimal hyperparameters for document image binarization has received little attention. Gatos \textit{et al.} \cite{gatos2006adaptive} proposed a parameter-free binarization approach that depends on detailed modelling of the document image background. Dawoud \cite{dawoud2007iterative} proposed a method based on cross-section sequence that combines results at multiple threshold levels into a single binarization result. Badekas and Papamarkos \cite{badekas2008estimation} introduced an approach that performs binarization over a range of parameter values, and estimates the final parameter values by iteratively narrowing the parameter range. Howe \cite{howe2013document} proposed an interesting method that optimizes a global energy function based on the Laplacian image, and automatically estimates the best parameter settings. The method dynamically sets the regularization coefficient and Canny thresholds for each image by using a stability criterion on the resultant binarized image. Mesquita \textit{et al.} \cite {mesquita2015parameter} investigated a racing procedure based on a statistical approach, named I/F-Race, to fine-tune parameters for document image binarization. Cheriet \textit{et al.} \cite{cheriet2013learning} proposed a learning framework for automatic parameter optimization of the binarization methods, where optimal parameters are learned using support vector regression. However, the limitation of this work is the dependence on ground truth for parameter learning.

This work uses Bayesian optimization \cite{shahriari2016taking,jones2001taxonomy} to efficiently infer the optimal values of the control parameters. Bayesian optimization is a general approach for hyperparameter tuning that has shown excellent results across applications and disciplines \cite{snoek2012practical,shahriari2016taking}. 

\section{Proposed Binarization Technique}
\label{sec:methodology}

The overall pipeline of the proposed automatic document image binarization technique is presented in Fig. \ref{fig_framework} using an example image from H-DIBCO 2016 dataset. The binarization algorithm is discussed in detail as follows.

\subsection{Document Image Binarization}
\label{subsec:binarization}
Given a degraded document image, adaptive thresholding using median filters is first performed that separates the foreground text from the background. The output is a grayscale image with reduced background noise and distortion. The grayscale image is then passed through two band-pass filters separately for further background noise removal. A high frequency band-pass filter is used to separate the fine detailed text from the background, and a low frequency band-pass filter is used for masking that image in order to remove great parts of the noise. Finally, the noise reduced grayscale image is converted into a binary image using Kittler's minimum error thresholding algorithm \cite{kittler1986minimum}. 

Figure \ref{fig_framework} illustrates the overall binarization algorithm for a given degraded document image. The image output generated at each filtering step is presented for better understanding of the background noise removal algorithm. It can be observed from Fig. \ref{fig_framework} that the input document image is heavily degraded with stains, wrinkles and contrast variation. After performing adaptive median filtering, the image becomes less noisy, and is enhanced further using the two band-pass filtering approach. The final binarized image represents the document image with foreground text preserved and noisy background removed. However, the performance of the binarization algorithm depends upon six hyperparameter values that include two control parameters required by the adaptive median filter, namely the local threshold and the local window size; and four control parameters required by the band-pass filters, namely the mask size for blurring the text, a local threshold and two window size values for high frequency and low frequency band-pass filters. The value of these hyperparameters must be chosen such that quality metrics corresponding to the binarized image (e.g., F-measure, Peak Signal-To-Noise Ratio (PSNR), etc. \cite{gatos2009icdar}) are maximized (or error is minimized). This corresponds to an optimization problem that must be solved to arrive at the best combination of hyperparameters. For example, the following optimization problem finds optimal values of $d$ hyperparameters ${\bf x} = (x_1, x_2, ..., x_d)$ with respect to maximizing the F-measure \cite{gatos2009icdar},
\begin{equation*}
\begin{aligned}
& \underset{\bf x}{\text{maximize}}
& & \mathrm{f_{measure}}(\bf x) \\
& \text{subject to}
& & min\_x_1 \leq x_1 \leq max\_x_1, \; min\_x_2 \leq x_1 \leq max\_x_2, \\
&&& \cdots, \; min\_x_d \leq x_d \leq max\_x_d,
\end{aligned}
\end{equation*}
where $min\_x_i$ and $max\_x_i$ span the search space for parameter $x_i$. No-reference image quality metrics \cite{mittal2012no} can be optimized in place of metrics such as F-measure in applications where ground truth reference images are not available.  

Techniques like grid search, cross-validation, evolutionary optimization, etc. can be used to find optimal values of the hyperparameters in the $d$-dimensional search space. For large values of $d$, such approaches tend to be slow. Bayesian optimization \cite{shahriari2016taking} efficiently solves such hyperparameter optimization problems as it aims to minimize the number of evaluations of the objective function (e.g., F-measure in this case) required to infer the hyperparameters, and is used in this work in the interest of computational efficiency. 



\subsection{Bayesian Optimization}
\label{subsec:bayesopt}
Bayesian optimization is a model based approach involving learning the relationship between the hyperparameters and the objective function \cite{jones2001taxonomy}. A model is constructed using a training set known as an \emph{initial design} that covers the hyperparameter space in a space-filling manner. Statistical designs such as Latin hypercubes and factorial designs \cite{singh2016design} can be used for this purpose. The initial design may also be random. The goal is to gather initial information about the parameter space in absence of any prior knowledge. 

After the model is trained using the initial design, an iterative sampling approach follows. The sampling algorithm uses the information offered by the model to intelligently select additional samples in regions of the hyperparameter space that are likely to lead to the optimal solution (i.e., good binarization performance as quantified by metrics such as F-measure). The model is then refined by extending the training set with the selected samples. Bayesian optimization involves using a Bayesian model such as Gaussian process models \cite{rasmussen2006gaussian}, and the sampling scheme exploits the mean and variance of prediction offered by the model to select additional samples iteratively. 
This work considers Gaussian process models in the context of Bayesian optimization. A detailed explanation of Gaussian processes can be found in \cite{rasmussen2006gaussian}.

\subsection{Gaussian Processes}
A Gaussian process (GP) is the multivariate Gaussian distribution generalized to an infinite-dimensional stochastic process where any finite combination of dimensions are jointly-Gaussian \cite{rasmussen2006gaussian}. A Gaussian process $f$ is completely described by its mean $m$ and covariance $k$ functions, $f({\bf x}) \sim \mathcal{GP}(m({\bf x}), k({\bf x, x'}))$. 

The mean function incorporates prior domain-specific knowledge, if available. The mean function $m({\bf x})=0$ is a popular choice without loss of generality in absence of any prior knowledge about the problem at hand. The covariance function $k$ incorporates variation of the process from the mean function and essentially controls the expressive capability of the model.

Numerous covariance functions exist in literature including squared exponential (SE) function \cite{snoek2012practical} and the Mat\'{e}rn  kernels \cite{rasmussen2006gaussian}. The SE kernel is a good general-purpose kernal and is used for experiments in this paper. The SE kernel is described as, $k({\bf x}_i, {\bf x}_j) = exp\Big( -\frac{1}{2\theta^2} \|{\bf x}_i - {\bf x}_j\|^2 \Big)$, with $\theta$ being a hyperparameter that controls the width of the kernel.

\noindent Let ${\mathcal{D}}=(X,{\bf y})$ be a $n$-point training set. Let $\bf K$ be the kernel matrix holding the pairwise covariances between points in $X$,
\begin{equation}
{\bf K} = 
\begin{bmatrix}
k({\bf x}_1, {\bf x}_1) & \dots  & k({\bf x}_1, {\bf x}_n) \\
\vdots & \ddots & \vdots \\
k({\bf x}_n, {\bf x}_1) & \dots  & k({\bf x}_n, {\bf x}_n)
\end{bmatrix}.
\end{equation}
Let $y_{n+1} = \hat{y}({\bf x}_{n+1})$ be the predicted value of a query sample ${\bf x}_{n+1}$ using the GP model $\hat{y}$. Since ${\bf y}$ and $y_{n+1}$ are jointly-Gaussian by definition of Gaussian processes,
\begin{equation}
\begin{bmatrix}
{\bf y} \\
y_{n+1}
\end{bmatrix}
\sim \mathcal{N}\Big( 0, 
\begin{bmatrix}
{\bf K} & {\bf k}\\
{\bf k}^\intercal & k({\bf x}_{n+1}, {\bf x}_{n+1})
\end{bmatrix}\Big),
\end{equation}
with ${\bf k} = [k({\bf x}_{n+1}, {\bf x}_1,), \dots k({\bf x}_{n+1}, {\bf x}_n,)]$. The posterior distribution is calculated as $P(y_{n+1}|{\mathcal{T}},{\bf x}_{n+1}) = \mathcal{N}(\mu_n({\bf x}_{n+1}), \sigma^2_n({\bf x}_{n+1}))$, where,
\begin{align}
\mu_n({\bf x}_{n+1}) &= {\bf k}^\intercal {\bf K}^{-1} {\bf y},\\
\sigma^2_n({\bf x}_{n+1}) &= k({\bf x}_{n+1}, {\bf x}_{n+1}) - {\bf k}^\intercal {\bf K}^{-1}{\bf k}.
\end{align}
The mean and variance of prediction is used by the sampling process to solve the optimization problem, and described as follows.

\begin{table*}
\begin{center}
\caption{Evaluation results of popular binarization methods on DIBCO datasets. The $^\ast$ marks the cases where existing binarization methods outperform the proposed approach.}
\label{tab:res1}
\resizebox{13.2cm}{!} { 
\begin{tabular}{|ll|c|c|c|c|c|}
\hline
Datasets & Methods & F-measure (\%)($\uparrow$) & PSNR ($\uparrow$) & DRD ($\downarrow$) & NRM (x $10^{-2}$) ($\downarrow$) & MPM (x $10^{-3}$) ($\downarrow$) \\ \hline
\multirow{8}{*}{DIBCO-2009} & Otsu \cite{otsu1975threshold} & 78.72 & 15.34 & N/A & 5.77 & 13.30 \\
 & Sauvola \cite{sauvola2000adaptive} & 85.41 & 16.39 & N/A & 6.94 & 3.20 \\
 & Niblack \cite{niblack1985introduction} & 55.82 & 9.89 & N/A & 16.40 & 61.50 \\
 & Bernsen \cite{bernse1986dynamic} & 52.48 & 8.89 & N/A & 14.29 & 113.80 \\
 & Gatos et al. \cite{gatos2006adaptive} & 85.25 & 16.50 & N/A & 10.00 & 0.70$^\ast$ \\
 & LMM \cite{su2010binarization} & 91.06$^\ast$ & 18.50$^\ast$ & N/A & 7.00 & 0.30$^\ast$ \\
 & Lu et al. \cite{lu2010document} & 91.24$^\ast$ & 18.66$^\ast$ & N/A & 4.31$^\ast$ & 0.55$^\ast$ \\
 & \textbf{Proposed method} & \textbf{90.58} & \textbf{18.13} & - & \textbf{5.50} & \textbf{2.26} \\
\hline
\multirow{8}{*}{H-DIBCO 2010} & Otsu \cite{otsu1975threshold} & 85.27 & 17.51 & N/A & 9.77 & 1.35 \\
 & Sauvola \cite{sauvola2000adaptive} & 75.3 & 15.96 & N/A & 16.31 & 1.96 \\
 & Niblack \cite{niblack1985introduction} & 74.10 & 15.73 & N/A & 19.06 & 1.06 \\
 & Bernsen \cite{bernse1986dynamic} & 41.30 & 8.57 & N/A & 21.18 & 115.98 \\
 & Gatos et al. \cite{gatos2006adaptive} & 71.99 & 15.12 & N/A & 21.89 & 0.41$^\ast$ \\
 & LMM \cite{su2010binarization} & 85.49 & 17.83 & N/A & 11.46 & 0.37$^\ast$ \\
 & Lu et al. \cite{lu2010document} & 86.41 & 18.14 & N/A & 9.06 & 1.11 \\
 & \textbf{Proposed method} & \textbf{89.65} & \textbf{18.78} & - & \textbf{5.82} & \textbf{0.66} \\
\hline
\multirow{11}{*}{DIBCO-2011} & Otsu \cite{otsu1975threshold} & 82.22 & 15.77 & 8.72 & N/A & 15.64 \\
 & Sauvola \cite{sauvola2000adaptive} & 82.54 & 15.78 & 8.09 & N/A & 9.20 \\
 & Niblack \cite{niblack1985introduction} & 68.52 & 12.76 & 28.31 & N/A & 26.38 \\
 & Bernsen \cite{bernse1986dynamic} & 47.28 & 7.92 & 82.28 & N/A & 136.54 \\
 & Gatos et al. \cite{gatos2006adaptive} & 82.11 & 16.04 & 5.42 & N/A & 7.13 \\
 & LMM \cite{su2010binarization} & 85.56 & 16.75 & 6.02 & N/A & 6.42 \\
 & Lu et al. \cite{lu2010document} & 81.67 & 15.59 & 11.24 & N/A & 11.40 \\
 & Lelore \cite{lelore2011super} & 80.86 & 16.13 & 104.48 & N/A & 64.43 \\
 & Howe \cite{howe2011laplacian} & 88.74$^\ast$ & 17.84$^\ast$ & 5.37 & N/A & 8.64 \\
 & Su et al. \cite{su2013robust} & 87.8 & 17.56$^\ast$ & 4.84 & N/A & 5.17 \\
 & \textbf{Proposed method} & \textbf{88.61} & \textbf{17.54} & \textbf{3.92} & - & \textbf{4.39} \\
\hline
\multirow{8}{*}{H-DIBCO 2012} & Otsu \cite{otsu1975threshold} & 80.18 & 15.03 & 26.45 & N/A & N/A \\
 & Sauvola \cite{sauvola2000adaptive} & 82.89 & 16.71 & 6.59 & N/A & N/A \\
 & LMM \cite{su2010binarization} & 91.54$^\ast$ & 20.14$^\ast$ & 3.05 & N/A & N/A \\
 & Improved Lu et al. \cite{lu2010document} & 90.38 & 19.30 & 3.35 & N/A & N/A \\
 & Su et al. \cite{su2012learning} & 87.01 & 18.26 & 4.42 & N/A & N/A \\
 & Lelore \cite{lelore2011super} & 92.85$^\ast$ & 20.57$^\ast$ & 2.66$^\ast$ & N/A & N/A \\
 & Howe \cite{howe2011laplacian} & 89.47 & 21.80$^\ast$ & 3.44 & N/A & N/A \\
 & \textbf{Proposed method} & \textbf{90.96} & \textbf{19.44} & \textbf{2.96} & - & - \\
\hline
\multirow{8}{*}{DIBCO-2013} & Otsu \cite{otsu1975threshold} & 83.94 & 16.63 & 10.98 & N/A & N/A \\
 & Sauvola \cite{sauvola2000adaptive} & 85.02 & 16.94 & 7.58 & N/A & N/A \\
 & LMM \cite{su2010binarization} & 92.12$^\ast$ & 20.68$^\ast$ & 3.10 & N/A & N/A \\
 & Howe \cite{howe2013document} & 92.70$^\ast$ & 21.29$^\ast$ & 3.18 & N/A & N/A \\
 & Combined \cite{howe2013document} and \cite{moghaddam2013unsupervised} & 91.81$^\ast$ & 20.68$^\ast$ & 4.02 & N/A & N/A \\
 & Combined \cite{moghaddam2013unsupervised} and \cite{nafchi2012historical} & 89.79 & 18.99 & 4.24 & N/A & N/A \\
 & Combined \cite{moghaddam2013unsupervised} and \cite{moghaddam2010multi} & 84.90 & 17.04 & 8.25 & N/A & N/A \\ 
 & \textbf{Proposed method} & \textbf{91.28} & \textbf{19.65} & \textbf{2.77} & - & - \\
\hline
\multirow{8}{*}{H-DIBCO 2014} & Otsu \cite{otsu1975threshold} & 91.78 & 18.72 & 2.65 & N/A & N/A \\
 & Sauvola \cite{sauvola2000adaptive} & 86.83 & 17.63 & 4.89 & N/A & N/A \\
 & Howe \cite{howe2013document} & 96.63$^\ast$ & 22.40$^\ast$ & 1.00$^\ast$ & N/A & N/A \\
 & Combined \cite{howe2013document} and \cite{mesquita2014new} & 96.88$^\ast$ & 22.66$^\ast$ & 0.90$^\ast$ & N/A & N/A \\
 & Modified \cite{nafchi2012historical} & 93.35 & 19.45 & 2.19 & N/A & N/A \\
 & Golestan University team \cite{ntirogiannis2014icfhr2014} & 89.24 & 18.49 & 4.50 & N/A & N/A \\
 & University of Thrace team \cite{ntirogiannis2014icfhr2014} & 89.77 & 18.46 & 4.22 & N/A & N/A \\
 & \textbf{Proposed method} & \textbf{93.79} & \textbf{19.74} & \textbf{1.90} & - & - \\
\hline
\end{tabular}
}
\end{center}
\end{table*}

\subsection{Sampling Algorithms}
A sampling scheme must make a trade-off between \emph{exploration} of the hyperparameter space, and \emph{exploitation} of sub-spaces with a high likelihood of containing the optima. The variance estimates provided by the GP offer insight on unexplored regions, while the mean predictions point towards estimates of the behavior of the objective function in a region of interest. Therefore, the model can be an effective tool to select a set of samples from a large number of candidates (e.g., generated randomly), that either enrich the model itself (by sampling unexplored regions), or drive the search towards the optima (by exploiting regions with optimal predicted objective function values).

Popular sampling schemes in literature include the expected improvement criterion, the probability of improvement criterion and upper/lower confidence bounds \cite{shahriari2016taking} (UCB/LCB). Sampling algorithms are also known as acquisition functions in the context of Bayesian optimization \cite{shahriari2016taking}.
The upper and lower confidence bounds offer a good mix of exploration and exploitation. The probability of improvement favors exploitation much more than exploration, while expected improvement lies in between the two. This work uses the UCB criterion for its balanced sampling characteristics in absence of any problem-specific knowledge. Let $\mu({\bf x})$ and $\sigma({\bf x})$ be the posterior mean and variance of prediction provided by the Gaussian process (GP) model. The value of the UCB criterion corresponding to a set of hyperparameters $\bf x$ is defined as, $\alpha_{UCB} ({\bf x}) = \mu(\bf x) + \beta \sigma ({\bf x})$.
This essentially corresponds to \emph{exploring} $\beta$ intervals of standard deviation around the posterior mean provided by the GP model. The value of $\beta$ can be set to achieve optimal regret according to well-defined guidelines \cite{srinivas2012information}.
A detailed discussion of the sampling approaches is out of scope of this work, and the reader is referred to \cite{shahriari2016taking,snoek2012practical} for a deeper treatment of sampling algorithms, and Bayesian optimization in general.

\section{Experiments}
\label{sec:experiments}
The following section demonstrates the proposed approach on benchmark datasets and compares it to existing approaches. 

\subsection{Experimental setup}
\label{sec:setup}
The proposed binarization method has been tested on the images from the DIBCO dataset \cite{gatos2009icdar,gatos2011icdar2,pratikakis2013icdar} that consists of machine-printed and handwritten images with associated ground truth available for validation and testing, and the H-DIBCO \cite{pratikakis2010h,pratikakis2012icfhr,ntirogiannis2014icfhr2014,pratikakis2016icfhr2016} dataset that consists of handwritten document test images. The performance of the proposed method is compared with the state-of-the-art binarization methods such as \cite{otsu1975threshold,sauvola2000adaptive,niblack1985introduction,bernse1986dynamic,su2010binarization,howe2013document}. Six hyperparameters of the binarization algorithm are automatically selected using Bayesian optimization. These include a local threshold $\tau_1$ and local window size $ws$ for adaptive median filtering; and local threshold $\tau_2$, mask size for blurring the text $ms$, window size $ws_h$ and $ws_l$ for high frequency and low frequency band-pass filters respectively. The corresponding optimization problem is formulated as,
\begin{equation*}
\begin{aligned}
& \underset{\bf x}{\text{maximize}}
& & \mathrm{f_{measure}}(\bf x) \\
& \text{subject to}
& & 0.05 \leq \tau_1 \leq 0.2, \; 35 \leq ws \leq 95, \; 0.05 \leq \tau_2 \leq 0.5, \\
&&& 0 \leq ms \leq 10, \; 200 \leq ws_h \leq 400, \; 50 \leq ws_l \leq 150,
\end{aligned}
\end{equation*}
where ${\bf x} = (\tau_1, ws, \tau_2, ms, ws_h, ws_l)$.
This work uses the Bayesian optimization framework available as part of MATLAB (R2017a). The parameter $\beta$ of UCB criterion was set to $2$.

\begin{figure}
\includegraphics[width=3.4in]{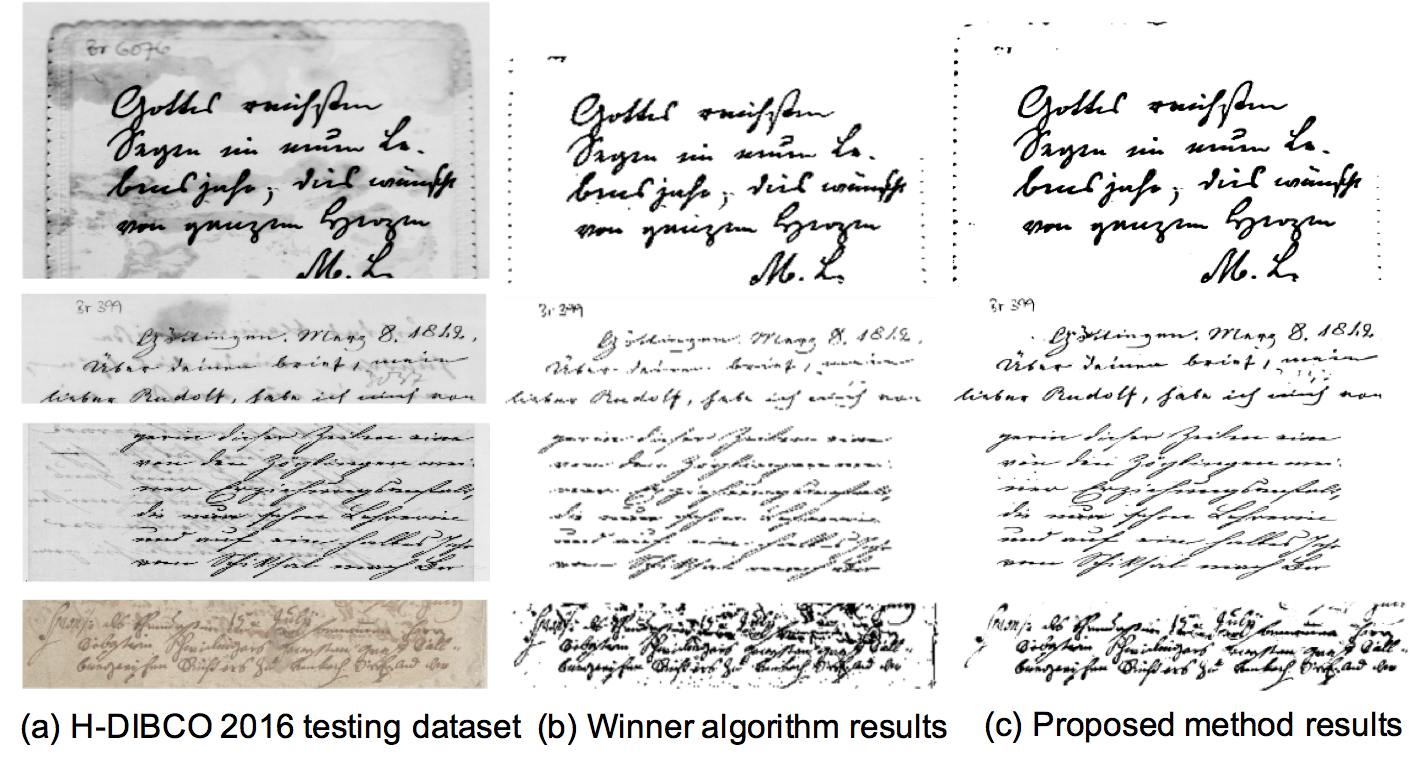}
\caption{Document image binarization results obtained on sample test images from the H-DIBCO 2016 dataset.}
\label{fig_binRes}
\end{figure}

\subsection{Experimental results}
\label{sec:results}

The evaluation measures are adapted from the DIBCO reports \cite{gatos2009icdar,pratikakis2010h,gatos2011icdar2,pratikakis2012icfhr,pratikakis2013icdar,ntirogiannis2014icfhr2014,pratikakis2016icfhr2016}, and include F-measure, Peak Signal-to-Noise Ratio (PSNR), Distance Reciprocal Distortion metric (DRD), Negative Rate Metric (NRM) and Misclassification Penalty Metric (MPM). The binarized image quality is better with high F-measure and PSNR values, and low DRD, MPM and NRM values. For details on the evaluation measures, the reader is referred to \cite{gatos2009icdar,pratikakis2016icfhr2016}.

\begin{table}
\begin{center}
\caption{Evaluation results on the H-DIBCO 2016 dataset and comparison with top ranked methods from the competition.}
\label{tab:res2}
\resizebox{8.3cm}{!} { 
\begin{tabular}{|c|l|c|c|c|}
\hline
Rank & Methods & F-measure (\%)($\uparrow$) & PSNR ($\uparrow$) & DRD ($\downarrow$) \\ \hline
1 & Technion team \cite{katz2007direct} & 87.61$\pm$6.99 & 18.11$\pm$4.27 & 5.21$\pm$5.28 \\
2 & Combined \cite{hassaine2011efficient} and \cite {hassaine2012set} & 88.72$\pm$4.68 & 18.45$\pm$3.41 & 3.86$\pm$1.57 \\
3 & Method based on \cite {hassaine2012set} & 88.47$\pm$4.45 & 18.29$\pm$3.35 & 3.93$\pm$1.37 \\
4 & UFPE Brazil team \cite{pratikakis2016icfhr2016} & 87.97$\pm$5.17 & 18.00$\pm$3.68 & 4.49$\pm$2.65 \\
5 & Method adapted from \cite{hassaine2011efficient} & 88.22$\pm$4.80 & 18.22$\pm$3.41 & 4.01$\pm$1.49 \\
- & Otsu \cite{otsu1975threshold} & 86.61$\pm$7.26 & 17.80$\pm$4.51 & 5.56$\pm$4.44\\
- & Sauvola \cite{sauvola2000adaptive} & 82.52$\pm$9.65 & 16.42$\pm$2.87 & 7.49$\pm$3.97 \\
- & \textbf{Proposed method} & \textbf{92.03$\pm$7.61} & \textbf{19.75$\pm$4.36} & \textbf{3.19$\pm$2.17}\\
\hline
\end{tabular}
}
\end{center}
\end{table}

\begin{table*}
\begin{center}
\caption{Comparison results of average F-measure (\%), PSNR and DRD values obtained using different binarization methods.}
\label{tab:res3}
\resizebox{13.2cm}{!} { 
\begin{tabular}{|l|c|c|c|c|c|c|c|}
\hline
\multicolumn{1}{|l|}{Methods}
& \multicolumn{2}{|c|}{DIBCO 2009-2016}
& \multicolumn{2}{|c|}{DIBCO 2009-2013} 
& \multicolumn{3}{|c|}{DIBCO 2011-2014}\\ \cline{2-8}
& \multicolumn{1}{|c|}{F-measure (\%) ($\uparrow$)}
& \multicolumn{1}{|c|}{PSNR ($\uparrow$)} 
& \multicolumn{1}{|c|}{F-measure (\%) ($\uparrow$)}
& \multicolumn{1}{|c|}{PSNR ($\uparrow$)} 
& \multicolumn{1}{|c|}{F-measure (\%) ($\uparrow$)}
& \multicolumn{1}{|c|}{PSNR ($\uparrow$)}
& \multicolumn{1}{|c|}{DRD ($\downarrow$)} \\
\hline
 Otsu \cite{otsu1975threshold} & 84.10 & 16.68 & 82.06 & 16.05 & 84.53 & 16.53 & 12.20\\ 
 Sauvola \cite{sauvola2000adaptive} & 82.93 & 16.54 & 82.23 & 16.35 & 84.32 & 16.76 & 6.78\\ 
 LMM \cite{su2010binarization} & - & - & 89.15 & 18.78$^\ast$ & - & - & -\\  
 Howe \cite{howe2013document} & - & - & - & - & 91.88$^\ast$ & 20.83$^\ast$ & 3.24\\  
 \textbf{Proposed method} & \textbf{90.99} & \textbf{19.00} & \textbf{90.21} & \textbf{18.71} & \textbf{91.16} & \textbf{19.09} & \textbf{2.88}\\ 
 \hline
\end{tabular}
}
\end{center}
\end{table*}

\begin{figure}
\centering
\includegraphics[width=2.3in]{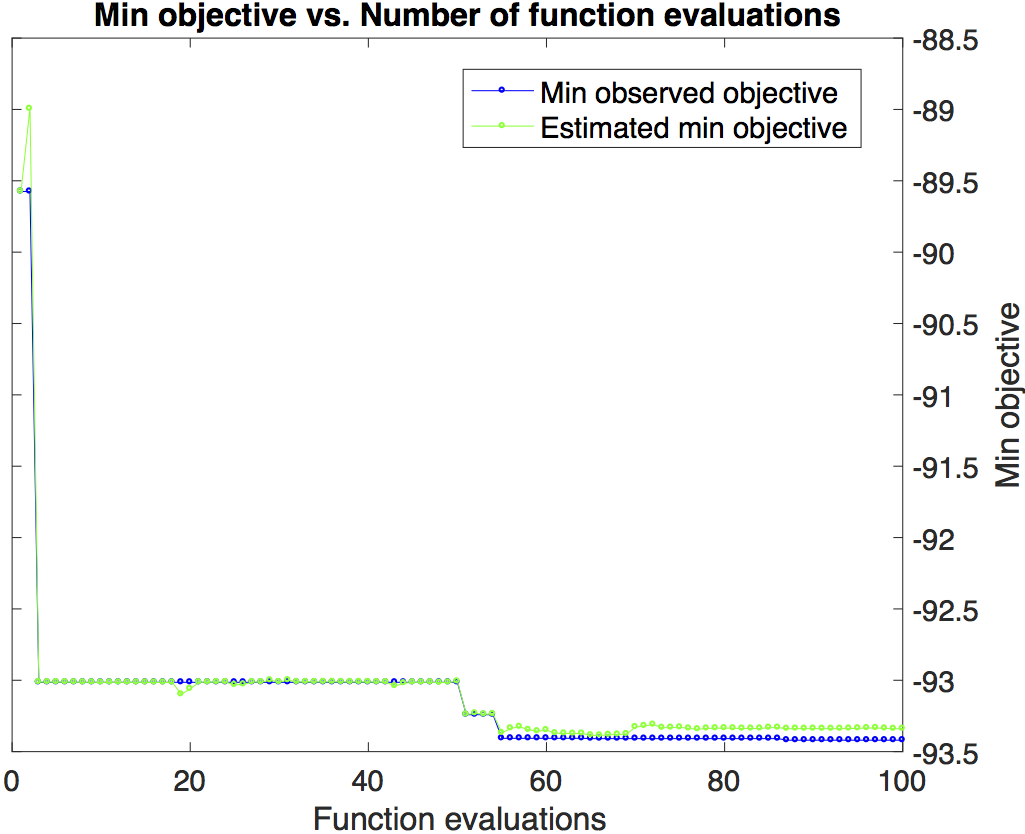}
\caption{The evolution of F-measure during the Bayesian optimization process. The estimated objective refers to the value of F-measure predicted by the GP model trained as part of the optimization process. The model accurately tracks the value of the objective function. The objective values are negative since the implementation followed the convention of minimizing the objective function rather than maximizing. Therefore, the objective function here is $-1 * F-Measure$. Figure best viewed in color.}
\label{fig_boConvergence}
\end{figure}

The experimental results are presented in Table \ref{tab:res1}. The proposed method is compared to several popular binarization methods on competition datasets from 2009 to 2014. Table \ref{tab:res2} illustrates the evaluation results on the most recent H-DIBCO 2016 dataset, and a comparison is drawn with the top five ranked methods from the competition, and the state-of-the-art methods. Figure \ref{fig_binRes} highlights the document binarization results for sample test images from the H-DIBCO 2016 dataset and compares with the results obtained from the algorithm of the competition winner. Finally, Table \ref{tab:res3} presents the average F-measure, PSNR and DRD values across different dataset combinations. In Tables \ref{tab:res1}-\ref{tab:res3}, $\uparrow$ implies that a higher value of F-measure and PSNR is desirable, while $\downarrow$ implies a lower value of DRD, NRM and MPM is desirable. The $^\ast$ indicates a case where the result of an existing method is better than the proposed method.

It is observed from Table \ref{tab:res2} and Figure \ref{fig_binRes} that the proposed method achieves higher scores with respect to F-measure, PSNR and DRD, as compared to other methods. However, on closely inspecting Table \ref{tab:res1}, it can be seen that there are instances where existing methods outperform the proposed method by a close margin (marked as $^\ast$). Nevertheless, with reference to all datasets used in the experiments, the proposed method is found to be most consistent and stable with high F-measure and PSNR, and low DRD, NRM and MPM scores. Table \ref{tab:res3} empirically evaluates the performance of the proposed method with respect to all 86 images from DIBCO 2009-2016. On an average, the proposed method achieves $90.99\%$ F-measure and $19.00$ PSNR for all test images under the experimental settings. For DIBCO 2009-2013, the top ranked method \cite{su2010binarization} from the competition achieves $89.15\%$ F-measure, and the proposed method outperforms it by achieving $90.21\%$ accuracy. The top ranked method \cite{howe2013document} in DIBCO 2011-2014 competition obtains $91.88\%$ accuracy, which is marginally higher (by $0.72\%$) than the accuracy achieved using the proposed method ($91.16\%$). The proposed method produces least visual distortions (DRD) in comparison to other methods.

Figure \ref{fig_boConvergence} conveys the accuracy of the GP model trained as part of the Bayesian optimization process. The estimated values of the F-measure (the green curve) are in line with the observed values (obtained by computing F-measure values of the selected samples, represented by the blue curve). This validates the accuracy of the GP model and subsequently, the correctness of the Bayesian optimization process. In general, the Bayesian optimization-based approach used herein can aid in automating state-of-the-art binarization methods.

\section{Conclusions}
\label{sec:conclusions}
A novel binarization technique is presented in this paper that efficiently segments the foreground text from heavily degraded document images. The proposed technique is simple, robust and fully automated using Bayesian optimization for on-the-fly hyperparameter selection. The experimental results on challenging DIBCO and H-DIBCO datasets demonstrate the effectiveness of the proposed method. On an average, the accuracy of the proposed method for all test images is found to be $90.99\%$ (F-measure). As future work, the ideas presented herein will be scaled to perform preprocessing of images in word spotting algorithms, and hybridization of the proposed technique with existing state-of-the-art binarization methods will be explored. 

\section*{Acknowledgment}
This work was supported by the Swedish strategic research programme eSSENCE, the Riksbankens Jubileumsfond (Dnr NHS14-2068:1), and the G\"oran Gustafsson foundation. 



\bibliographystyle{IEEEtran}
\bibliography{refs}
%

\end{document}